\documentclass[psfig]{aa}

\usepackage{graphics}

\newcommand{\FIG}[1]{#1}

\newcommand{\PREP}[1]{}

\newcommand{\GG}{\mbox{\bf g}}
\newcommand{\II}{\mbox{\bf I}}
\newcommand{\ee}{\hat{{\bf e}}_r}
\newcommand{\BB}{\mbox{\bf B}}
\newcommand{\vv}{\mbox{\bf v}}

\begin{document}

\thesaurus{7(02.13.2; 03.13.4; 06.19.2; 08.23.3)}

\title{Numerical simulations of stellar winds: polytropic models.}

\author{R.~Keppens \and J.P.~Goedbloed}

\institute{FOM-Institute for Plasma-Physics Rijnhuizen, P.O. Box 1207,
          3430 BE Nieuwegein, The Netherlands, \\
          keppens@rijnh.nl jpg@rijnh.nl}

\offprints{R.~Keppens}

\date{Received 8 september 1998 / Accepted}

\titlerunning{Numerical simulations of stellar winds}

\maketitle

\begin{abstract}
We discuss steady-state transonic outflows obtained by direct
numerical solution of the hydrodynamic and magnetohydrodynamic
equations. We make use of the Versatile Advection Code, a software
package for solving systems of (hyperbolic)
partial differential equations. We proceed
stepwise from a spherically symmetric, isothermal, unmagnetized, 
non-rotating Parker wind to arrive at axisymmetric, polytropic,
magnetized, rotating models. These represent 2D generalisations of
the analytical 1D Weber-Davis wind solution, which we obtain in the process.
Axisymmetric wind solutions containing both a `wind' and a `dead' zone
are presented.

Since we are solving for steady-state solutions, we efficiently exploit fully
implicit time stepping. The method allows us to
model thermally and/or magneto-centrifugally driven stellar outflows.
We particularly emphasize the \linebreak
boundary conditions imposed at the
stellar surface. For these axisymmetric, steady-state solutions, we can 
use the knowledge of the flux functions to verify the physical correctness of
the numerical solutions.
\end{abstract}


\section{Introduction}\label{s-intro}

Observational and theoretical research on stellar winds
and astrophysical jets has evolved rapidly. For our own sun and its
associated solar wind, the current understanding necessitates the
combined study of solar wind acceleration and coronal heating in 
time-dependent modeling (Holzer \& Leer \cite{holzer}). 
At the same time, Holzer and
Leer rightfully stressed that it remains
useful to emphasize on early studies of wind acceleration. This, we find, is
especially true for numerical modeling of stellar winds. With the ultimate
goal of time-dependent heating/wind modeling in mind, we here address the
simpler question on how to 
accurately model 1D and 2D steady-state winds by the numerical
solution of the polytropic magnetohydrodynamic (MHD) equations.

Since much of the 1D solutions we obtain is known from the outset, we can
verify our results {\it precisely}. Indeed, 1D Parker (\cite{parker})
and Weber-Davis (Weber \& Davis \cite{wd}) 
wind solutions can be checked to agree with
the analytic description. In their axisymmetric 2D extensions,
various physical quantities must be conserved along poloidal streamlines. 
Sakurai (\cite{sakuraiAA,sakurai}) 
presented such 2D generalizations of the magnetized
Weber-Davis wind, using a method designed from these conservation laws. 
With modern numerical schemes, we can recover and extend his solutions
to allow for the self-consistent modeling of `dead' and `wind' zones,
as in the solar wind. Our steady-state solutions can be checked to conserve
quantities along streamlines {\it a posteriori}. 

Of crucial importance is the choice of boundary conditions used
in the simulations. Since the governing equations for steady-state,
transonic MHD flows are of mixed-type, their character can
change from elliptic to hyperbolic at a priori undetermined
internal critical surfaces. Causality arguments have been used to discuss
which and how many boundary conditions must be prescribed 
(Bogovalov \cite{bogovAA}). Our
choice of boundary conditions used at the stellar surface
is therefore discussed in detail.

All solutions presented are obtained with a single software package, the
Versatile Advection Code (VAC, see T\'oth~\cite{vacapjl,vac2}
and also {\tt http://www.phys.uu.nl/}\~{\tt toth}).
Although we only present steady-state transonic outflows
in spherical and axisymmetry, VAC is
developed for handling hydrodynamic (HD) and 
magnetohydrodynamic (MHD) one-, two-, or three-dimensional,
steady-state or time-de\-pen\-dent problems in astrophysics. It is therefore
capable of achieving our ultimate goal of time-dependent 3D wind modeling.
The insight gained in this study of steady-state polytropic flows
will then be very useful.

In Sect.~\ref{s-eq}, we list the equations and discuss the
Versatile Advection Code for solving them in Sect.~\ref{s-vac}.
Our calculations are presented in Sects.~\ref{s-1dwind},~\ref{s-2dhd},
and~\ref{s-2dmhd}. Conclusions are given in Sect.~\ref{s-concl}.
The approach taken is a gradual one, where for instance our 1D solutions 
are used to construct initial conditions for their 2D extension. We will
therefore model, in increasing order of complexity: (i) isothermal,
spherically symmetric Parker winds; (ii) polytropic, spherically symmetric
Parker winds; (iii) polytropic, rotating Parker winds for the equatorial
plane; (iv) Weber-Davis magnetized, polytropic, rotating winds for
the equatorial plane; and finally axisymmetric, polytropic, rotating 2D winds,
both (v) unmagnetized and magnetized, without (vi) and with (vii) a `dead'
zone. 

\section{Equations}\label{s-eq}

We solve the HD and MHD equations expressed in
the conservative variables density $\rho$, momentum vector $\rho \vv$,
and magnetic field $\BB$. These are given by
\begin{equation}
\frac{\partial \rho}{\partial t}+\nabla \cdot (\rho \vv)=0,
\label{q-mass}
\end{equation}
\begin{equation}
\frac{\partial (\rho \vv)}{\partial t}+ \nabla \cdot [ \rho \vv \vv + p_{tot} \II-
\BB \BB]=\rho \GG,
\label{q-mom}
\end{equation}
\begin{equation}
\frac{\partial \BB}{\partial t}+ \nabla \cdot (\vv \BB-\BB \vv)= 0.
\label{q-b}
\end{equation}
We introduced $p_{tot}=p + \frac{1}{2}B^2$ as the total pressure,
$\II$ as the identity tensor, $\GG$ as the external gravitational field,
and exploited
magnetic units such that the magnetic permeability is unity.
We drop the energy equation and assume a
polytropic relation connecting the thermal pressure $p$ and
the density $\rho$. For a polytropic index $\gamma$, we
thus assume $p\sim\rho^{\gamma}$. Hence, we do not address
the heat deposition in the corona. 
Although we solve the time-dependent equations
as given above, we will only present steady-state $\partial/\partial t=0$
solutions of Eqs.~(\ref{q-mass})--(\ref{q-b}). 
For stellar wind calculations, we consider a spherically symmetric 
external gravitational field 
$\GG=-G M_*/r^2\ee$, where $G$ is the gravitational constant,
$M_*$ is the stellar mass, $r$ is
the distance to the stellar center, and $\ee$
indicates the radial unit vector.

\section{Versatile Advection Code}\label{s-vac}

In this section, we discuss the software and numerical method used. The 
physics results are described from Sect.~\ref{s-1dwind} onwards.
The Versatile Advection Code (VAC, see T\'oth~\cite{vacapjl,vac2}) 
is a general purpose software package
for solving a conservative system of hyperbolic partial differential
equations with additional non-hyperbolic source terms, such as the
MHD equations. VAC runs on PC's, on a variety of workstations, on
vector platforms, and we can also run in parallel on a cluster of
workstations, and on distributed memory architectures like
the Cray T3D and T3E, and the IBM SP (Keppens \& T\'oth~\cite{vac4},
T\'oth \& Keppens~\cite{vac3}).
The code is written in the dimension independent LASY syntax 
(T\'oth~\cite{lasy}),
so it can be used as a convenient tool 
to handle HD and MHD one-, two-, or three-dimensional problems in 
astrophysics and laboratory plasma physics.
The dimensionality of the problem and the actual set of equations to
solve are easily selected in a preprocessing step. 

VAC uses a structured finite volume grid and offers a choice of
conservative, second order accurate, shock-capturing,
spatial and temporal discretization schemes. The spatial discretizations
include two Flux Corrected Transport variants and four Total
Variation Diminishing (TVD) schemes (T\'oth \& Odstr\v cil~\cite{vac1}). 
Temporal discretization can
be explicit, semi-implicit, or fully implicit. It was recently
demonstrated (Keppens et al.~\cite{implvac1}, T\'oth et al.~\cite{implvac2}) 
how the implicit approach can be used very efficiently,
for steady-state and time-accurate problems possibly containing
discontinuities. Here, we expect smooth solutions to
the steady-state HD and MHD equations, so one can greatly
benefit computationally from fully implicit time stepping.

In this paper, we solve the polytropic HD ($\BB=0$) Eqs.~(\ref{q-mass})
and~(\ref{q-mom}) in 
Sects.~\ref{ss-parker} and~\ref{s-2dhd}. Magnetohydrodynamic
equations are solved in Sects.~\ref{ss-wd} and~\ref{s-2dmhd}. We
solve one-dimensional problems in Sect.~\ref{s-1dwind} and 
two-\-dimensional problems in Sects.~\ref{s-2dhd} and~\ref{s-2dmhd}. 
In practice, this means that the stellar winds we model are
solutions of the equations under the additional
assumption of a prescribed symmetry in the ignored directions. 
One-dimensional problems assume a spherical symmetry, while 2D solutions
assume $\partial/\partial \varphi=0$. Here, $\varphi$ denotes the
angle in a cylindrical $(R,\varphi,z)$
coordinate system centered on the star with its polar and
rotation axis as $z$-axis.

Since we are interested in steady-state solutions,
we use fully implicit time stepping as
detailed and demonstrated in Keppens et al. (\cite{implvac1}) and 
T\'oth et al. (\cite{implvac2}).
The linear systems arising in the linearized fully implicit backward
Euler scheme are solved using a direct block tridiagonal solver for
the 1D problems and using a preconditioned Stabilized Bi-Conjugate Gradient
iterative solver (van der Vorst \cite{henk}) for the 2D cases. 
The Modified Block Incomplete $LU$ preconditioner
is described in van der Ploeg et al.~(\cite{auke}).
We consistently used the TVD Lax-Friedrich (TVDLF) 
spatial discretization (Yee~\cite{yee}, T\'oth \& Odstr\v cil \cite{vac1})
using {\it Woodward} limiting (Collela \& Woodward~\cite{wood}). 
We typically took Courant numbers $C={\cal O}(100)$. For all 1D solutions
and for the 2D hydrodynamic solutions, the steady-state is
reached when the relative change in the conservative variables from one time
level to the next drops below $10^{-8}$. We use a
normalized measure defined by (T\'oth et al.~\cite{implvac2})
\begin{equation}
  \Delta_2 U \equiv \sqrt{
    \frac{1}{N_\mathrm{var}}\sum_{u=1}^{N_\mathrm{var}}
    \frac{\sum_\mathrm{grid} (U_u^{n+1}- U_u^n)^2}{\sum_\mathrm{grid} (U^n_u)^2},
}
\end{equation}
where $N_{var}$ is the number of conserved variables $U_i$, and the
superscripts indicate the pseudo time level $t^n$.
Since we are solving for smooth solutions, the numerical schemes
can easily achieve such accuracy in the steady-state solutions.

For the axisymmetric 2D MHD solutions, the way to ensure a
zero divergence of the magnetic field is non-trivial. Unless one uses a
scheme which keeps $\nabla \cdot \BB =0$ exactly in some discretization,
like the constrained transport method (Evans \& Hawley~\cite{evans}), 
corrective action needs to be taken
during the time integration. This involves either including corrective
source terms in the equations which are proportionate to the numerically
generated divergence (Powell~\cite{powel}, their use for TVDLF was first
advocated by T\'oth \& Odstr\v cil~\cite{vac1}), or making use of a projection
scheme (Brackbill \& Barnes~\cite{barnes}) which involves the solution of
a Poisson equation. These two approaches can even be combined and
such combination is also beneficial for 
fully implicit schemes (T\'oth et al.~\cite{implvac2}). For the 2D axisymmetric
MHD wind solutions presented in Sect.~\ref{s-2dmhd}, it proved difficult
to ensure a divergence free solution in a fully implicit manner. Using
explicit time stepping and employing the projection scheme before every
time step, we could get steady-state solutions where 
$\Delta_2 U \simeq {\cal O}(10^{-7})$
which have acceptable $\mid \nabla \cdot \BB \mid < 10^{-3}$, although
this time-marching method is computationally much more costly than
the implicit approach. The use of Powell source terms alone proved inadequate
for these wind solutions, as fairly large errors
are then advected into the whole solution domain from
the stellar boundary outwards. 

VAC makes use of two layers of ghost cells surrounding the
physical domain to implement boundary conditions. A symmetry condition
at a boundary is then imposed by mirroring the calculated values in
the two physical cell layers adjacent to the boundary in these ghost cells.
The boundary conditions imposed at the stellar surface are extremely important.
We will emphasize them for all cases considered.

\section{1D polytropic winds}\label{s-1dwind}

\subsection{Parker winds}\label{ss-parker}

Our starting point is the well-known analytic Parker (\cite{parker}) 
solution for
a spherically symmetric, isothermal ($\gamma=1$) outflow from a
star of mass $M_*$ and radius $r_*$. Given the
magnitude of the escape speed $v_{esc}=\sqrt{2GM_*/r_*}$,
one can construct a unique
`wind' solution which starts subsonically at the stellar surface 
and accelerates monotonically to supersonic speeds.
This solution is transonic at the critical position 
$r_s=v_{esc}^2 r_*/(4c_{s*}^2)$,
where $c_{s*}^2=p/\rho$ is the constant isothermal sound speed.
Since we know the position of the critical point $r_s$, we can easily
determine the flow profile $v_r(r)$ and the corresponding density
profile $\rho(r)$. The radial velocity is obtained from the iterative
solution of the transcendental equation
\begin{equation}
\frac{v_r}{c_{s*}}=\sqrt{4\left(\frac{r_s}{r}\right)-3+2\ln
     \left[\left(\frac{r}{r_s}\right)^2\left(\frac{v_r}{c_{s*}}\right)\right]}.
\end{equation}
The density profile results from the integrated mass conservation equation.
Since the radial velocity reaches a constant supersonic value asymptotically,
the corresponding density vanishes at infinity as $1/r^2$.

Choosing units such that $r_*=1$, $\rho_*\equiv\rho(r_*)=1$ with $c_{s*}=1$, 
we initialize a 1D spherically
symmetric, polytropic (with $\gamma>1$) hydrodynamic outflow with this
analytic isothermal Parker wind solution on a non-uniform mesh ranging
through $r\in [1,50]r_*$. We use 1000 grid points and exploit
a grid accumulation at the stellar surface,
where the acceleration due to the pressure gradient is largest.
In the ghost cells used to impose
boundary conditions at the stellar surface, we fix the value of
the base density to unity, and extrapolate the radial momentum continuously 
from its calculated value in the first grid cell. At $r=50 r_*$, we extrapolate
both density and radial momentum continuously into the ghost cells.
We then use a fully implicit time integration to arrive at the corresponding
steady-state, spherically symmetric, polytropic Parker wind solution.
The obtained solution $\rho(r)$, $v_r(r)$
can be verified to have a constant mass flux
$\rho r^2 v_r$ as a function of radius and energy integral
\[E=\left(v_r^2/2+c_s^2/(\gamma-1)-GM_*/r\right)/c_{s*}^2,\]
 where $c_s^2(r)=\rho^{\gamma-1}$.
Also, determining the sonic point $r_s$ where $v_r(r_s)=c_s(r_s)$ and
the base radial velocity $v_{r*}$, the solution can be checked to satisfy
\begin{equation}
\frac{r_s}{r_*}=\left(\frac{v_{esc}}{2 c_{s*}}\right)^{\frac{\displaystyle 2(\gamma+1)}{\displaystyle 5-3\gamma}}\left(\frac{c_{s*}}{v_{r*}}\right)^{\frac{\displaystyle 2(\gamma-1)}{\displaystyle 5-3\gamma}}.
\label{q-checkpoly}
\end{equation}
Note how the isothermal $\gamma=1$ case is the only polytropic wind
solution where the position of the critical point can be determined
{\it a priori}.

\begin{figure}
\begin{center}
\FIG{
\resizebox{\columnwidth}{!}{\includegraphics{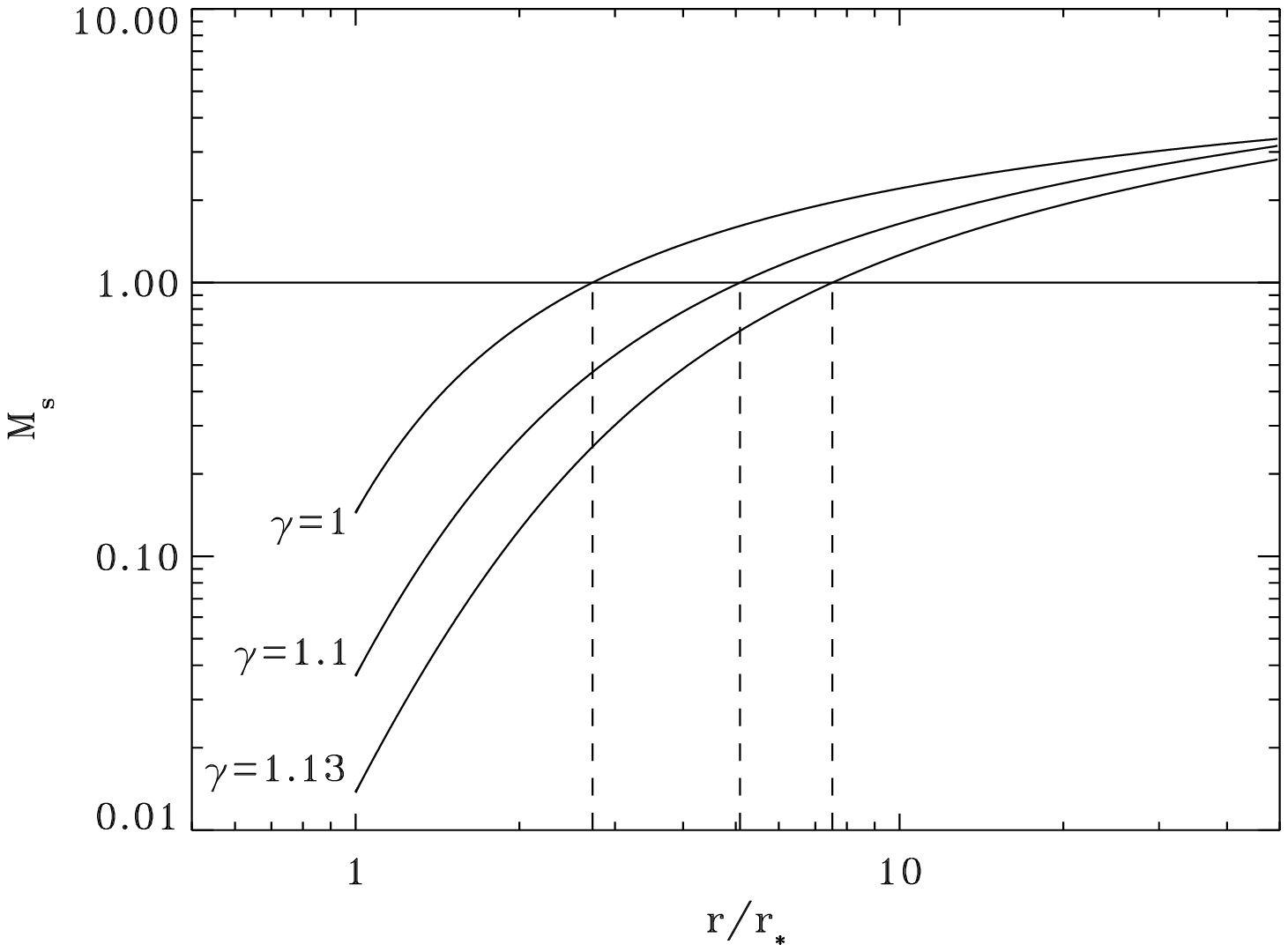}} \\
\resizebox{\columnwidth}{!}{\includegraphics{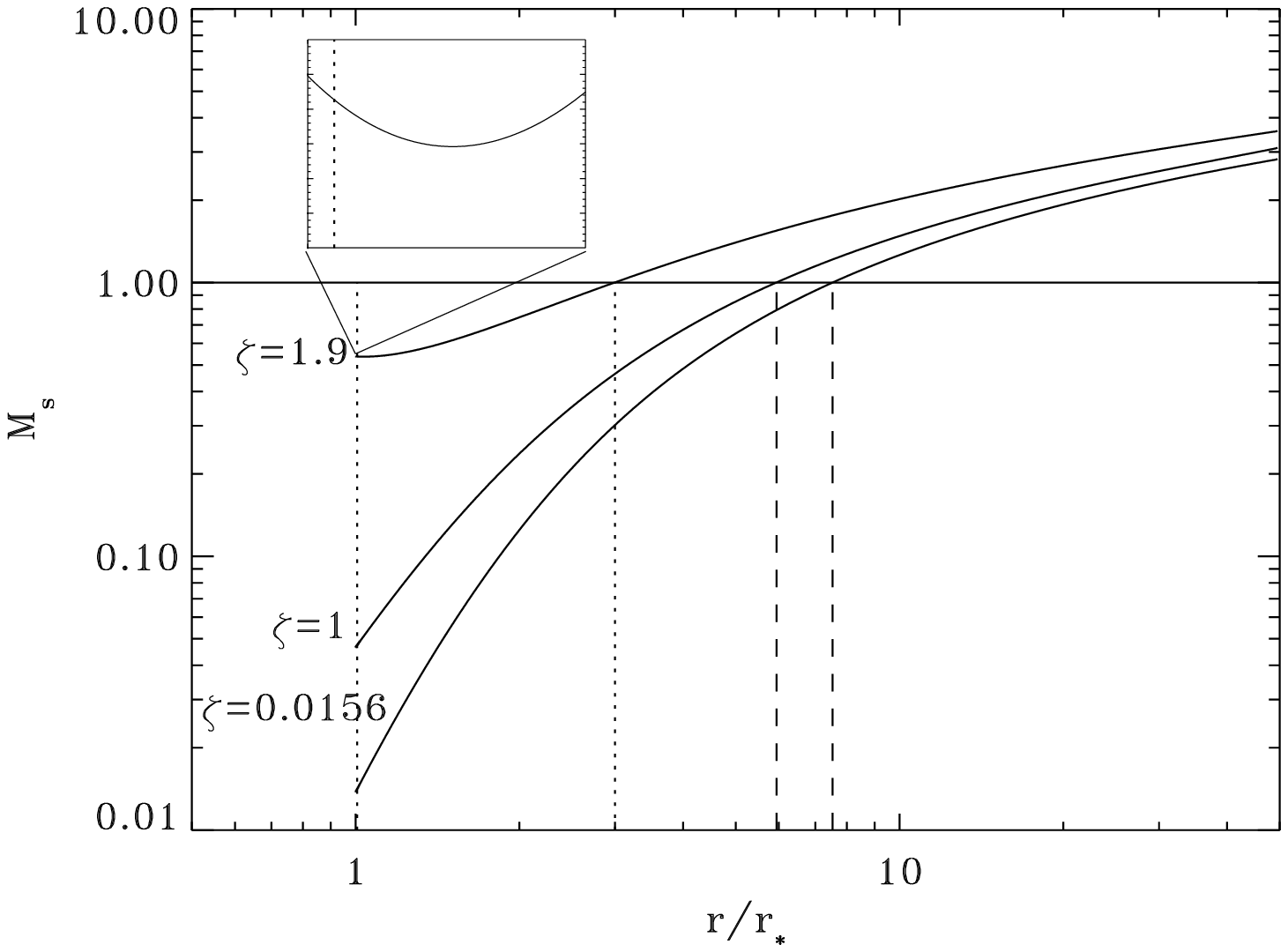}}
}
\caption{Polytropic Parker winds with $v_{esc}=3.3015 c_{s*}$. 
Top panel: Mach number $M_s$ as a function
of radial coordinate for spherically symmetric
Parker winds for polytropic index $\gamma=1$ (isothermal), $\gamma=1.1$
and $\gamma=1.13$. Bottom panel: equatorial solutions for rotating
$\gamma=1.13$ polytropic Parker winds for various rotation parameters
$\zeta$. Critical points are indicated.}
\label{f-parpoly}
\end{center}
\end{figure}

In practice, we increased the polytropic index gradually from
$\gamma=1$ through 1.05, 1.1, 1.12, 1.125 to $\gamma=1.13$, each time
relaxing the obtained steady-state solution for one polytropic index
to the unique transonic wind solution for the next value. In the top panel
of
Fig.~\ref{f-parpoly}, we plot the radial variation of the Mach number
$M_s=v_r/c_s$ for the isothermal $\gamma=1$ Parker wind with
$v_{esc}=3.3015 c_{s*}$, and
for similar polytropic winds with $\gamma=1.1$ and $1.13$. The vertical
dashed lines indicate the
agreement of the positions of the sonic points where $M_s=1$ with
the calculated right hand side of Eq.~(\ref{q-checkpoly}). 
Note the outward shift of the sonic point with
increasing polytropic index and the corresponding decrease 
of the asymptotic radial velocity.

When we relax the restriction of spherical symmetry by allowing a rigid
stellar rotation rate $\Omega_*$, we can easily construct a solution
for the equatorial plane only. Indeed, ignoring variations
perpendicular to this plane, one simply adds a toroidal velocity profile
where $v_{\varphi}(r)=\Omega_* r_* (r_*/r)$ and then solves
for $\rho(r)$, $v_r(r)$, and $v_{\varphi}(r)$. The boundary conditions on the
toroidal momentum keep $\rho v_{\varphi}$ fixed at the base and extrapolate
it continuously
at the end of the computational domain. A polytropic,
rotating Parker solution for the equatorial regions is found by 
relaxation from a non-rotating wind with the same polytropic index 
$\gamma$. In the bottom panel of Fig.~\ref{f-parpoly}, we show the Mach number
$M_s(r)$ for $v_{esc}=3.3015 c_{s*}$ and $\gamma=1.13$ 
Parker winds where $\zeta=\Omega_* r_*/c_{s*}$ equals
$\zeta=0.0156$, $\zeta=1$ and $\zeta=1.9$.
The solution with $\zeta=0.0156$ hardly differs from its non-rotating
thermally driven counterpart shown in Fig.~\ref{f-parpoly}, as expected.
The additional centrifugal
acceleration causes an increase in the 
base velocity $v_{r*}$ and in the asymptotic
radial velocity.
Again, the solution can be verified to have a constant radial
mass flux $\rho r^2 v_r$, Bernoulli function 
\[E=\left([v_r^2+v_{\varphi}^2]/2+c_s^2/(\gamma-1)-GM_*/r\right)/c_{s*}^2,\] 
and constant specific angular momentum $r v_{\varphi}$. The positions of
the critical point(s) $r_s$ are now obtained from a generalization
of Eq.~(\ref{q-checkpoly}), namely from the solutions of
\begin{equation}
\zeta^2\left(\frac{r_*}{r_s}\right)^2-\frac{v_{esc}^2 r_*}{2 c_{s*}^2 r_s}
+2\left[\left(\frac{v_{r*}}{c_{s*}}\right)\left(\frac{r_*}{r_s}\right)^2\right]^\frac{\displaystyle 2(\gamma-1)}{\displaystyle \gamma+1}=0.
\label{q-parpolrot}
\end{equation}
This equation reduces to a second degree polynomial for a $\gamma=1$ isothermal,
rotating, Parker wind so it is evident that rotation rates exist
that introduce a second critical point. 
In Fig.~\ref{f-parpoly}, only the $\zeta=1.9$ solution exhibits
two critical points, shown as vertical dotted lines, within the domain. 
We determined the critical point(s) by solving Eq.~(\ref{q-parpolrot}) 
using the calculated base speed $v_{r*}$.
The close-up of the radial variation of $M_s$ for $\zeta=1.9$
at the base reveals that this thermo-centrifugally driven wind passes
the first critical point while being
decelerated, then starts to accelerate and finally becomes supersonic at
the second critical point. To correctly capture the dynamics close to the
stellar surface 
it is clear that we need a high grid resolution, 
especially at the stellar surface.
Indeed, the first critical point for the $\zeta=1.9$ solution 
is situated at $1.006 r_*$.

\subsection{Weber-Davis winds}\label{ss-wd}

The magnetized Weber-Davis (WD) solution (Weber \& Davis~\cite{wd}) represents a
valuable extension to the rotating, polytropic Parker
wind solution for the equatorial plane.
Again assuming that there is no variation perpendicular to
this plane, one now needs to solve for an additional two
magnetic field components $B_r(r)$ and $B_{\varphi}(r)$.
One is trivially obtained from the $\nabla \cdot \BB =0$ equation,
namely $B_r=B_{r*}r_*^2/r^2$. The analytic treatment reveals
that the magnetized polytropic wind solution has a total
of two critical points, namely
the slow $r_s$ and the fast $r_f$ critical point.
These are determined by
the zeros of $v_r^4-v_r^2(c_s^2+A_r^2+A_{\varphi}^2)+c_s^2 A_r^2$.
In between lies the Alfv\'en point $r_A$, defined as the radius at which
the radial velocity $v_r$ equals the radial Alfv\'en 
speed $A_r=B_r/\sqrt{\rho}$. Since the equatorial fieldline is prescribed
to be radial in the poloidal plane and the transfield force
balance is not taken into account, this Alfv\'enic transition is not a
critical point in this model.

\begin{figure}
\begin{center}
\FIG{
\resizebox{\columnwidth}{!}{\includegraphics{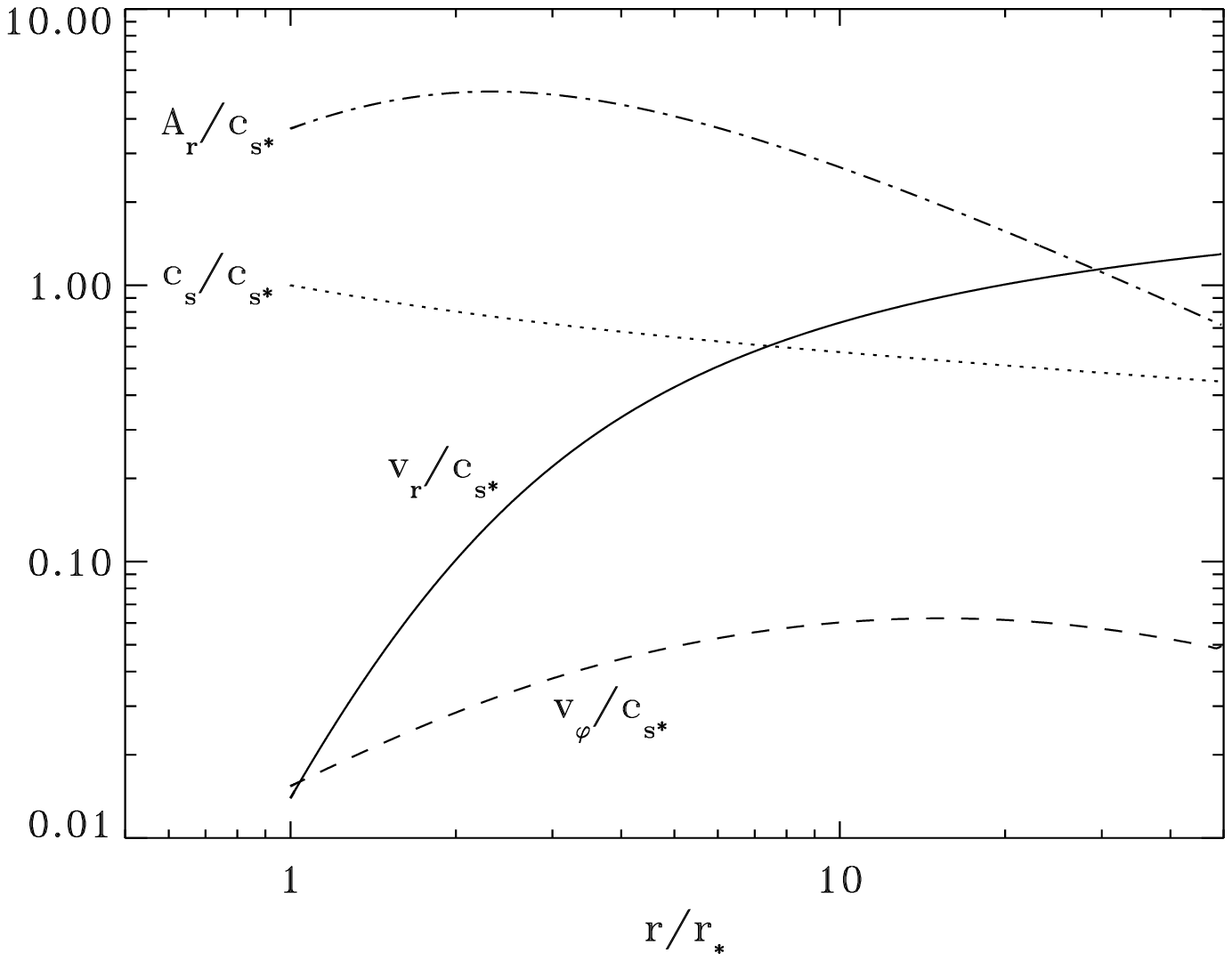}}
\resizebox{\columnwidth}{!}{\includegraphics{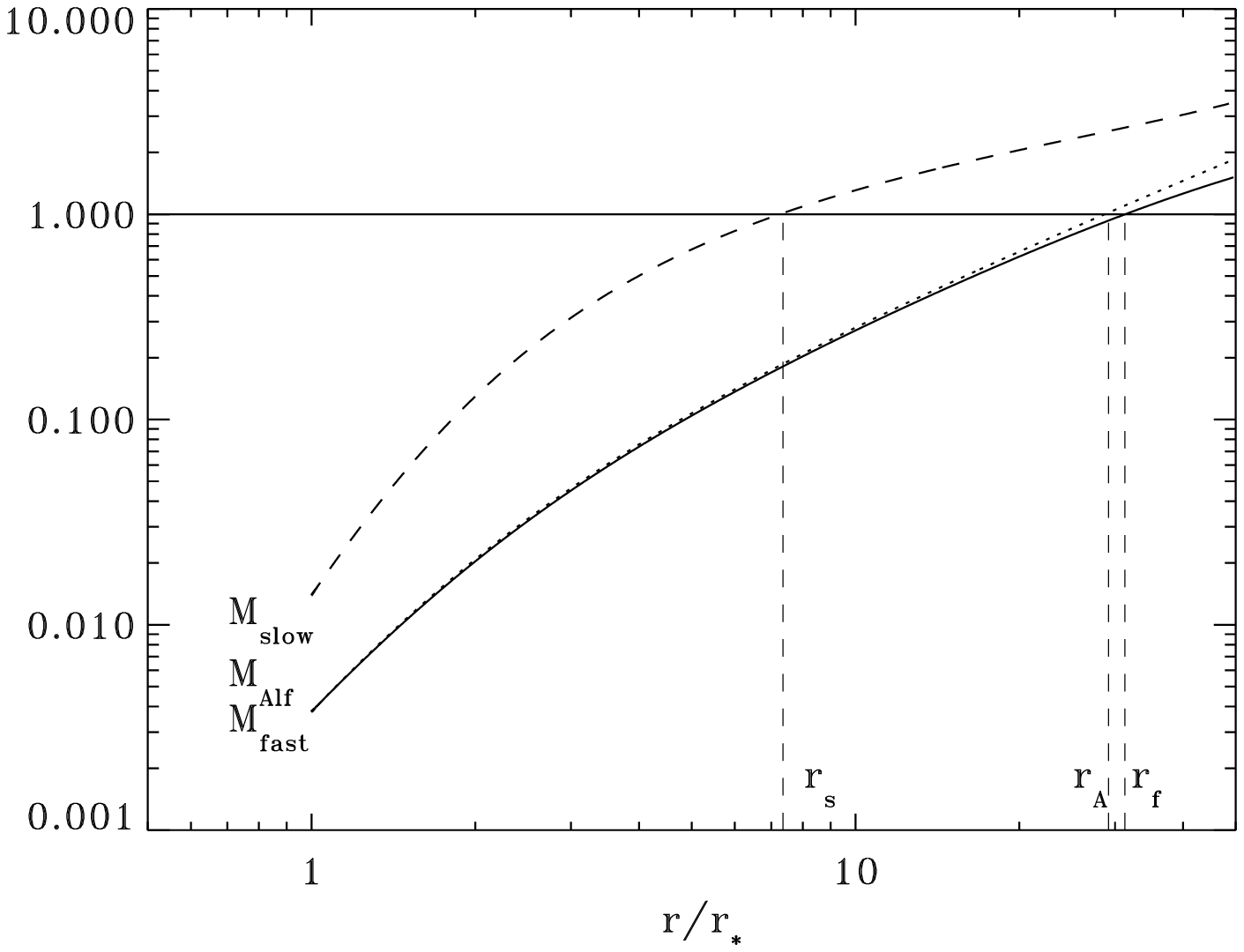}}
}
\caption{Weber-Davis wind solution for the equatorial plane.
Top panel: radial variation of the radial Alfv\'en speed $A_r$, sound
speed $c_s$, and velocities $v_r$ and $v_\varphi$,
all normalized to the base sound speed $c_{s*}$. 
Bottom panel: the
corresponding poloidal Alfv\'en Mach number $M_A=v_r/A_r$, and
poloidal slow and fast Mach numbers, determining the positions of the
critical points and the Alfv\'en point. See text for parameters.}
\label{f-wd}
\end{center}
\end{figure}

In the fully implicit time stepping towards a steady-state
WD wind for specific values
of $v_{esc}=3.3015 c_{s*}$, $\gamma=1.13$, $\zeta=0.0156$, and
for the base radial Alfv\'en speed $A_{r*}=B_{r*}/\sqrt{\rho_*}=3.69 c_{s*}$,
we initialize $\rho(r)$, $v_r(r)$, and $v_{\varphi}(r)$ with the
corresponding non-magnetic, polytropic rotating Parker solution.
We fix $B_r$ to its known $1/r^2$ dependence throughout the time
evolution, and initialize $B_{\varphi}$ to zero. The boundary conditions
at $r=50 r_*$ extrapolate all quantities we solve for continuously into
the ghost cells. At the base, we keep the density fixed, the radial momentum 
and toroidal field component are extrapolated linearly from the
first two calculated mesh points, while the toroidal momentum 
$\rho v_{\varphi}$ is coupled to the magnetic field ensuring
\begin{equation}
\frac{v_{\varphi}-\Omega_* r_*}{v_r}=\frac{B_{\varphi}}{B_r}.
\label{q-corotation}
\end{equation}
This expresses the parallelism of the velocity and the magnetic
field in the frame rotating with the stellar angular velocity $\Omega_*$.
Using these initial and boundary conditions, we arrive at the unique WD wind
solution for the given parameters. This magnetized polytropic wind
solution for the equatorial plane is shown in Fig.~\ref{f-wd}.  
The solution agrees {\it exactly} with the analytic WD wind: we obtain
five constants of motion, namely the mass flux $\rho r^2 v_r \simeq 0.0139$, 
the magnetic flux 
$r^2 B_r = 3.69$ which is constant by construction, 
the validity of Eq.~(\ref{q-corotation}) over the whole domain, 
the Bernoulli integral 
\[
\begin{array}{ccl}
E&=&\left([v_r^2+v_{\varphi}^2]/2+c_s^2/(\gamma-1)-GM_*/r
-v_\varphi B_\varphi B_r/\rho v_r \right. \\
 & & \left. + B_\varphi^2/\rho\right)/c_{s*}^2 \simeq 2.45, 
\end{array}
\]
and the constant total specific
angular momentum $L=r v_{\varphi}-r B_\varphi B_r/\rho v_r \simeq 13.36$.
The positions of the critical points are $r_s=7.4 r_*$ and $r_f=31.2 r_*$,
while the Alfv\'en point is at $r_A=29.2 r_*$,
as indicated in Fig.~\ref{f-wd}. This agrees with
the values given in the Appendix to Keppens et al. (\cite{rotpaper}), 
where the same
WD solution was calculated in a completely different fashion. Indeed,
the WD solution for given values of $\gamma$, $c_{s*}$, 
$v_{esc}/c_{s*}$, $\zeta$, and $A_{r*}/c_{s*}$, can alternatively be calculated
as a minimization problem in a six-dimensional space (see Belcher \&
MacGregor~\cite{belmac})
where one solves for the six unknowns 
[$v_{r*}$, $v_{\varphi *}$, $r_s$, $v_r(r_s)$, $r_f$, $v_r(r_f)$].
This can be done using standard Newton-Raphson iteration provided one
has an educated initial guess, but it can even be obtained by the
use of a {\it genetic algorithm}, as first demonstrated by 
Charbonneau (\cite{genetic}).
The fact that the, for our method initially unknown, base velocities appear 
in the determining set of variables for these minimization methods again
indicates that a high base resolution is absolutely essential. 
Values for these six unknowns found from
the solution in Fig.~\ref{f-wd} are $[0.01395,0.01541,7.4,0.6018,
\linebreak 31.2,1.1592]$
and these agree with the Newton-Raphson solution.

The calculation of the WD polytropic wind by the stepwise relaxation
from an isothermal Parker wind is thus an excellent test for
the numerics, as every step (from isothermal to
polytropic, from non-rotating to rotating, from Parker to WD) can
be verified {\it precisely} to agree with the
known solutions. It should be clear that we can construct WD wind solutions
where the acceleration results from the combined action of
thermal, centrifugal, and magnetic forces. However, our interest is in
the generalization of these 1D models out of the equatorial plane. 
We will again proceed in logical steps towards this goal.

\section{Axisymmetric 2D polytropic HD winds}\label{s-2dhd}

To arrive at a crude model for the coronal expansion
of a rigidly rotating star, we set forth to construct an
axisymmetric, steady-state,
polytropic wind solution valid throughout a poloidal cross-section.
With the polar axis as rotation and symmetry axis, 
we need to generalize the rotating, polytropic Parker wind 
which succesfully modeled the equatorial regions.
Whereas the Parker solution had at least one critical point, its 2D
extension is expected to give rise to critical curves in the poloidal
plane. The degree of rotation determines the deviation from perfect
circles arising in the non-rotating, spherically symmetric case.

To initialize a 2D fully implicit time-stepping procedure to arrive
at a steady-state wind, we use the 1D Parker solution with identical
escape speed $v_{esc}$, polytropic index $\gamma$, and rotational
parameter $\zeta$. We use a spherical $(r,\theta)$ grid in the poloidal
plane, where the grid spacing is equidistant in $\theta$, 
but is accumulated at the base in the radial direction. 
We take a $300 \times 20$ grid and only model a quarter of a full poloidal
cross-section.
The density is initialized such that for all angles $\theta$, the radial variation
equals the 1D Parker wind appropriate for the equator.
Writing the Parker solution as $\rho^P(r)$, $v_r^P(r)$, $v_{\varphi}^P(r)$, we
set $\rho(r,\theta; t=0)=\rho^P(r)$, and similarly, we set
$v_r(r,\theta; t=0)=v_r^P(r)$ and $v_{\varphi}(r,\theta; t=0)=v_{\varphi}^P(r)
\sin(\theta)$ so that it vanishes at the pole $\theta=0$, while
$v_{\theta}(t=0)=0$ everywhere. Since we now use a coarser radial resolution, 
we interpolate the Parker solution linearly onto the new radial grid.
Boundary conditions then impose symmetry conditions at
the pole ($\theta=0$) and the equator ($\theta=\pi/2$). 
The radial coordinate still covers
$r\in [1,50]r_*$, as in the 1D calculations. Since the
solutions are supersonic at $r=50 r_*$, the boundary conditions
there merely extrapolate the density and all three momentum components
linearly in the ghost cells. 
The stellar rotation enters as a boundary condition in the toroidal
momentum component, which enforces $v_{\varphi}=\Omega_* R_*$, where $(R,z)$ are
the cartesian coordinates in the poloidal $(r,\theta)$ plane.
Note that the toroidal momentum may still change in the process, since
we can no longer fix the density at the stellar surface to a $\theta$-independent
constant value. This is because in steady-state, the density profile should
establish a gradient in the $\theta$ direction to balance the component of
the centrifugal force in that direction. In the purely radial direction, the inwards
pointing gravity must be balanced by the combination of the
pressure gradient and the radial
component of the centrifugal force. We therefore extrapolate the density
linearly at the base.
To enforce the total mass flux as in the equatorial
Parker solution, we determine the constant $f_{mass}=\rho^P r^2 v_r^P$
from the 1D calculation, and
fix $\rho v_R=f_{mass}R/r^3$ and $\rho v_z=f_{mass}z/r^3$ 
at the stellar surface for its 2D extension. 

An elementary analytic treatment for a 2D polytropic steady-state
wind solution proceeds by noting that mass conservation is ensured
when the poloidal momentum is derived from an arbitrary stream
function $\chi(R,z)$ such that $\rho \vv_p = (1/R)\hat{e}_{\varphi}\times
\nabla \chi$. It is then easily shown that the toroidal momentum
equation is equivalent with the existence of a second arbitrary function
$L(\chi)=R v_{\varphi}$, corresponding to the conservation of
specific angular momentum along a poloidal streamline. 
Similarly, energy conservation along a streamline introduces 
\[E(\chi)=\left[v_R^2+v_z^2+v_{\varphi}^2\right]/2+
\rho^{\gamma -1}/(\gamma-1) - G M_*/r.\]
Across the poloidal streamlines
all forces must balance out.

\begin{figure}
\begin{center}
\FIG{
\resizebox{\columnwidth}{!}{\includegraphics{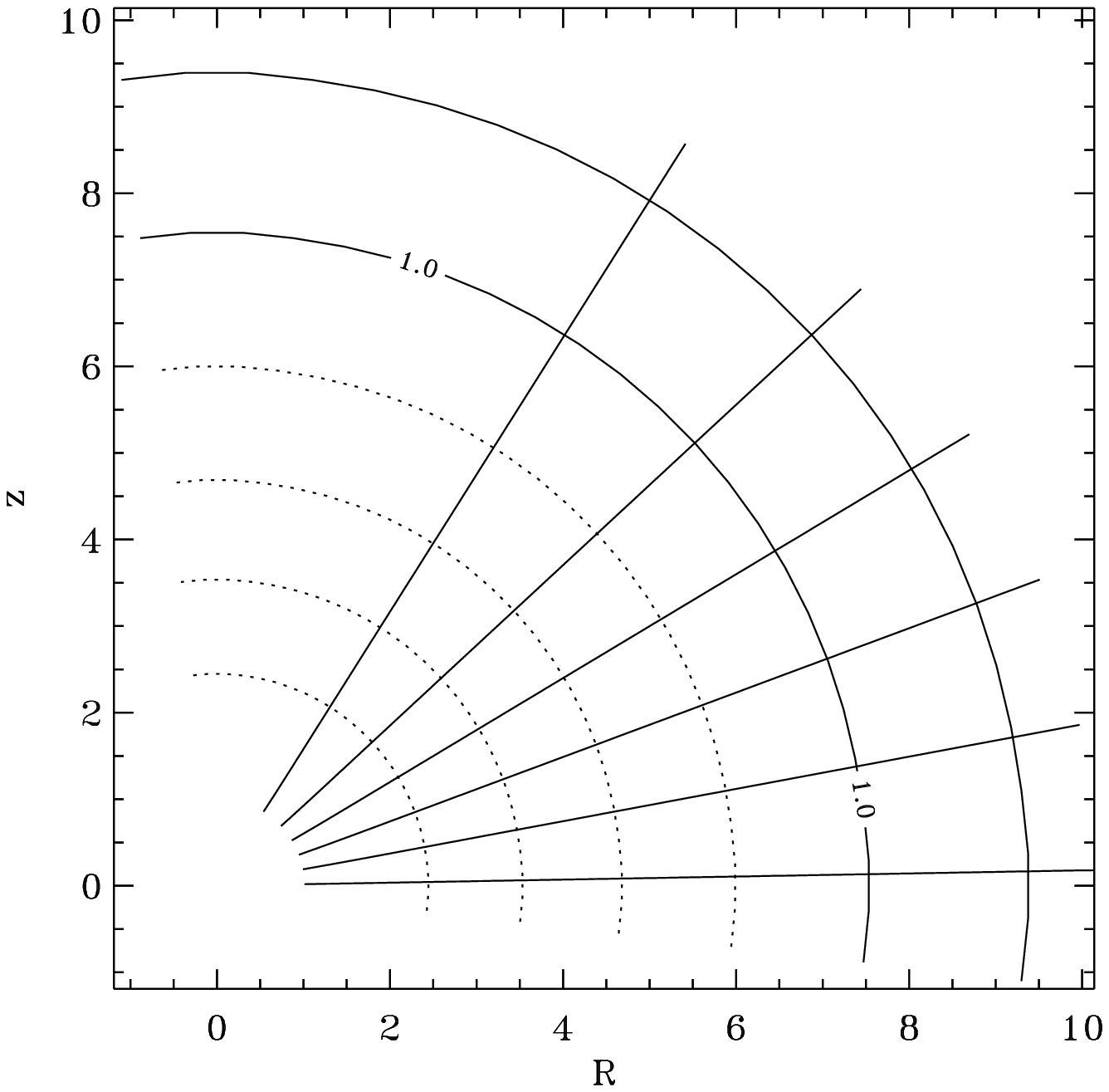}}
\resizebox{\columnwidth}{!}{\includegraphics{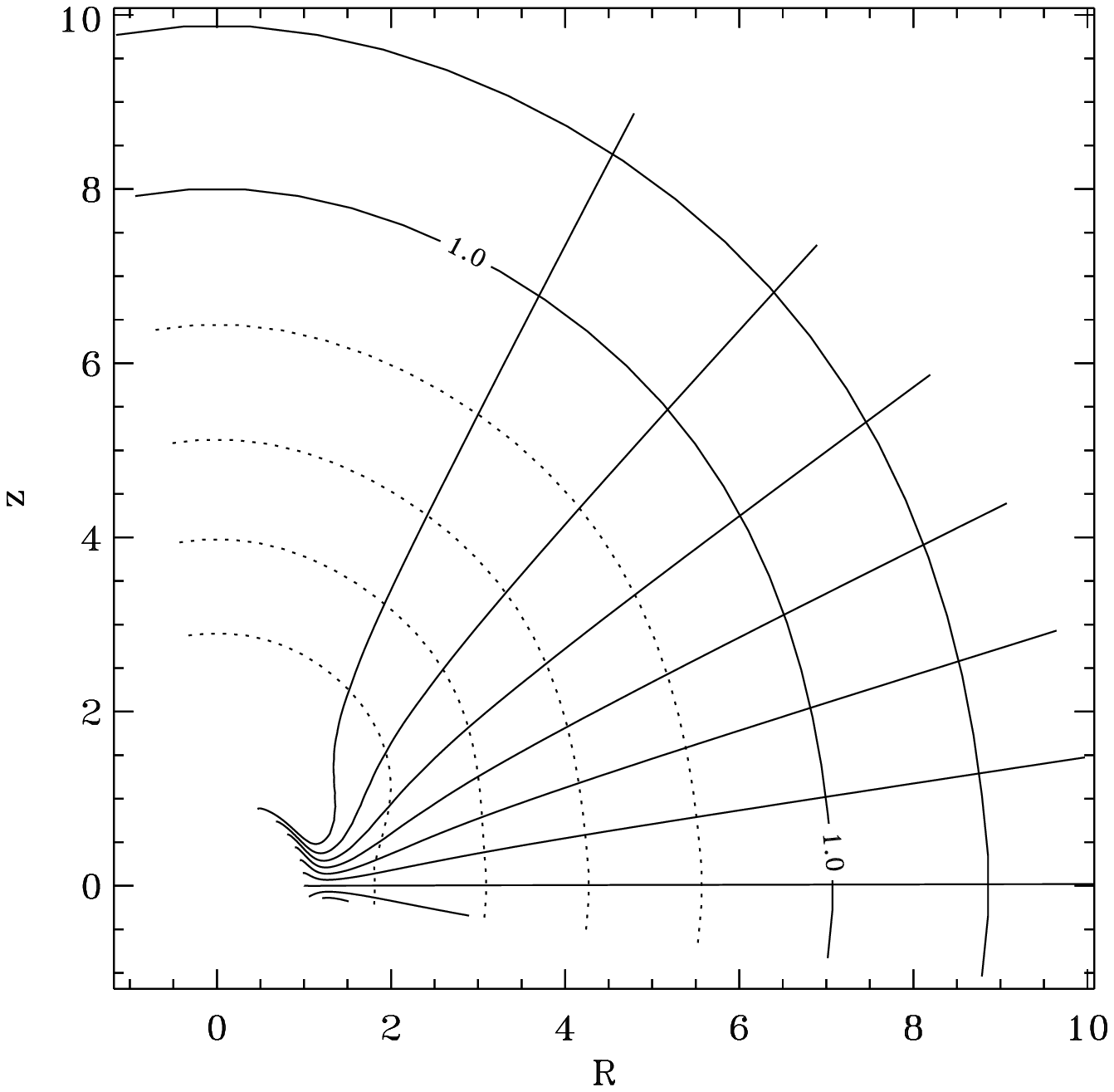}}
}
\caption{2D Polytropic HD winds. We show streamlines and contours 
(dotted for values below unity) of the
poloidal Mach number $M_p$ in the poloidal plane. 
For low (top panel) and high (bottom panel) rotation rates.}
\label{f-polyhd2d}
\end{center}
\end{figure}

We show streamlines and the contours of constant po\-loi\-dal Mach number
$M_p=\sqrt{(v_R^2+v_z^2)/c_s^2}$ for two hydrodynamic wind solutions
for $v_{esc}=3.3015 c_{s*}$, $\gamma=1.13$, and
with $\zeta=0.0156$ (top panel) and $\zeta=0.3$ (bottom panel)
in Fig.~\ref{f-polyhd2d}. We restricted the plotting region to about
$10 r_*$. For the imposed mass flux parameter, we used
the values $f_{mass}=0.01377$ for the slow rotator
and $f_{mass}=0.01553$ for the faster rotator, as found from the
equatorial Parker solution for the same $\gamma$ and $\zeta$. Note
how for low rotation rates, the wind solution is almost spherically
symmetric with nearly radial streamlines and circular
Mach curves. For higher rotation rates, the critical Mach curve where
$M_p=1$ moves inwards at the equator and outwards at the pole when
compared to a non-rotating case. The streamlines show the 
equatorward deflection when material is released
from the stellar surface due to the centrifugal force. 
Such equatorward streamline bending due to rotation is discussed in detail
in the analytical study by Tsinganos \& Sauty (\cite{tsinhd}).
For these solutions, we can then verify that the specific angular momentum
$L$, as well as the total energy $E$, are conserved along the streamlines.

\section{Axisymmetric 2D polytropic MHD winds}\label{s-2dmhd}

\subsection{Magnetized winds}

To obtain an axisymmetric magnetized wind solution, we may simply
add a purely radial magnetic field
to a 2D HD wind solution and use this configuration as an
initial condition for an MHD calculation. Hence, we set
$B_R(R,z; t=0)=\beta R/r^3$ and $B_z(R,z; t=0)=\beta z/r^3$ while
$B_\varphi(t=0)=0$. Such a monopolar field is rather unrealistic
for a real star, but it is the most straightforward way
to include magnetic effects. The same type of field was used
by Sakurai (\cite{sakuraiAA,sakurai}), 
which contained the first 2D generalization
of the WD model. 

Boundary conditions at equator
and pole are imposed by symmetry considerations. At $r=50 r_*$,
we extrapolate all quantities linearly. Similarly, the base conditions
extrapolate the density profile $\rho$ and all magnetic
field components, while the poloidal velocity components are
set to ensure a prescribed mass flux.
The stellar rotation rate and the coupling between the velocity
and the magnetic field enters in the boundary condition
at the stellar surface where we demand
\begin{equation}
v_\varphi=\Omega_* R_*+B_\varphi \sqrt{v_R^2+v_z^2}/\sqrt{B_R^2+B_z^2}.
\label{q-bc}
\end{equation}
Specific attention is paid to ensuring the $\nabla \cdot \BB=0$ condition.
As explained in Sect.~\ref{s-vac}, 
we now switch strategy and use explicit time stepping combined
with a projection scheme to obtain the steady state solution.

\begin{figure}
\begin{center}
\FIG{
\resizebox{\columnwidth}{!}{\includegraphics{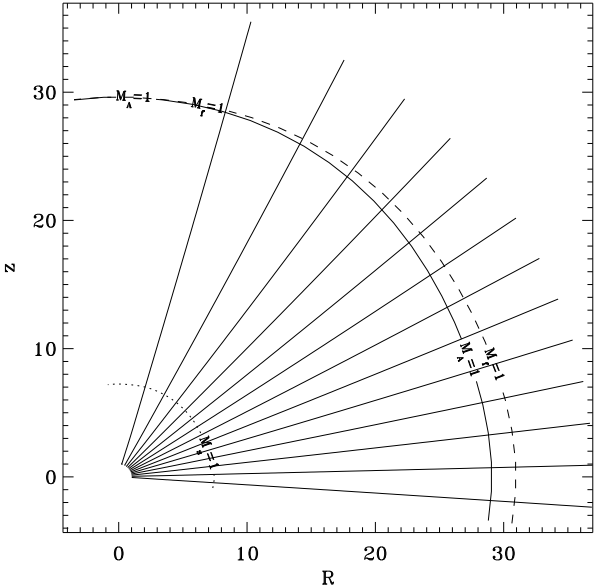}}
}
\caption{Polytropic axisymmetric MHD wind. We show the streamlines
and the positions of the critical surfaces where the poloidal
Mach numbers equal unity. Parameters are as in the 1D Weber-Davis wind
shown in Fig.~\ref{f-wd}.}
\label{f-polymhd2d}
\end{center}
\end{figure}

We calculated the 2D extension of the WD wind corresponding to
$v_{esc}=3.3015 c_{s*}$, $\gamma=1.13$, $\zeta=0.0156$ and
$\beta=3.69$. The mass flux is set to be $f_{mass}=0.01377$.
We show in Fig.~\ref{f-polymhd2d} the streamlines, and
the positions of the critical curves where the
poloidal Alfv\'en Mach number 
and the poloidal slow and fast
Mach numbers equal unity. 
The squared poloidal Alfv\'en Mach number $M_A^p$ is given by
\begin{equation}
\left(M_A^p\right)^2
=\left(v_R^2+v_z^2\right)/\left(A_R^2+A_z^2\right) \equiv \frac{v_p^2}{A_p^2}, 
\label{q-machAlfp}
\end{equation}
with Alfv\'en speeds $A_i\equiv B_i/\sqrt{\rho}$.
The squared poloidal slow $M_s^p$ and fast $M_f^p$ Mach numbers are defined by
\begin{eqnarray}
\left(M_s^p\right)^2 & = & \frac{2\left(v_R^2+v_z^2\right)}
{c_s^2+A_p^2+A_\varphi^2
-\sqrt{\left[c_s^2+A_p^2+A_\varphi^2\right]^2
       -4\,c_s^2\,A_p^2}}, \nonumber \\
 & &
\label{q-machslowp}
\end{eqnarray}
\begin{eqnarray}
\left(M_f^p\right)^2 & = & \frac{2\left(v_R^2+v_z^2\right)}
{c_s^2+A_p^2+A_\varphi^2
+\sqrt{\left[c_s^2+A_p^2+A_\varphi^2\right]^2
       -4\,c_s^2\,A_p^2}}. \nonumber \\
 & &
\label{q-machfastp}
\end{eqnarray}
At the pole, the fast Mach number coincides with the
Alfv\'en one since $B_\varphi$ vanishes there and the
parameters are such that $A_p^2>c_s^2$. Away from the pole,
the toroidal field component does not vanish, so that Alfv\'en
and fast critical curves separate. Note how the 
equatorial solution strongly resembles the WD wind
solution for the same parameters shown in Fig.~\ref{f-wd}.
The obtained wind solution is
mostly thermally driven, like the solar wind. The rotation rate and
magnetic field effects are minor and an almost spherically symmetric wind
results. Sakurai (\cite{sakuraiAA,sakurai}) 
demonstrated that for stronger fields, the
magnetic force of the spiraling fieldlines deflect the outflow poleward. This
magnetic pinching force can produce a polar collimation of the wind.
These effects have also been addressed by analytical studies of self-similar
outflows in Trussoni et al. (\cite{trusso}). 

For these axisymmetric, steady-state MHD outflows, the solutions can be
verified to obey the following conservation laws. Mass conservation
is ensured when writing the poloidal momentum vector as 
$\rho \vv_p = (1/R)\hat{e}_{\varphi}\times \nabla \chi$, with the stream 
function $\chi(R,z)$. The zero divergence of the magnetic field yields,
likewise, $\BB_p = (1/R)\hat{e}_{\varphi}\times \nabla \psi$, 
with $\psi$ the flux function. The poloidal part of
the induction equation then leads to $\chi(\psi)$, provided that the
toroidal component of the electric field $E_\varphi$ vanishes. This can
easily be checked from $v_R B_z=v_z B_R$, and the solution shown in
Fig.~\ref{f-polymhd2d} satisfies this equality to within 1\%. This
allows us to
write $\chi'\equiv d\chi/d\psi =\rho v_p/B_p=\rho v_R/B_R=\rho v_z/B_z$.  
A fair amount of algebra shows that the toroidal momentum and induction
equation introduce two more flux functions, namely the specific angular
momentum $L(\psi)=R v_\varphi -R B_\varphi B_p/\rho v_p$ and a function related
to the electric field 
$\Omega(\psi)=\left[v_\varphi-(v_p/B_p)B_\varphi\right]/R$.
The Bernoulli function derivable from the momentum equation can be
written as
\[
\begin{array}{ccl}
E(\chi)& =&\left[v_R^2+v_z^2+v_{\varphi}^2\right]/2+
\rho^{\gamma -1}/(\gamma-1) - G M_*/r  \\
 & & - v_\varphi B_\varphi B_p/\rho v_p + B_\varphi^2/\rho.
\end{array}
\]
Note how the constants of motion found in the WD solution
immediately generalize in this formalism 
(mass flux, magnetic flux, corotation as in Eq.~(\ref{q-corotation}),
specific angular momentum $L$ and Bernoulli function $E$). The hydrodynamic
limit is found for zero magnetic field $\BB=0$ and vanishing electric field
$\Omega=0$. Across the poloidal streamlines, 
the momentum balance is governed by the
generalized, mixed-type Grad-Shafranov equation. The numerical solutions
we obtained indeed have parallel poloidal streamlines and
poloidal fieldlines and conserve all these quantities along them.

\subsection{Winds containing a `dead' zone}

The monopolar field configuration used above is unrealistic. However, it
should be clear that
our method easily generalizes to bipolar stellar fields by
appropriately changing the initial condition on the magnetic field.
In fact, a star like our sun has open fieldlines at both poles and
closed fieldlines around its equator.
To obtain a steady-state stellar wind
containing a `wind' zone along the open fieldlines and a `dead' zone about
the stellar equator, we can simply initialize the polar regions up to
a desired polar angle $\theta_{wind}$ as above. The equatorial `dead' zone
is then initialized as follows: the density and the toroidal momentum
component is taken from the 2D HD wind
with the same rotational and polytropic parameters while the poloidal
momentum components are set to zero. The initial magnetic field configuration
in the `dead' zone is set to a dipole field which has
\begin{equation}
B_R=3 a_d \frac{z\,R}{(R^2+z^2)^{5/2}},
\end{equation}
and
\begin{equation}
B_z= a_d \frac{(2 z^2 -R^2)}{(R^2+z^2)^{5/2}}.
\end{equation}
The strength of the dipole is taken $a_d=\beta/(2 \cos(\theta_{wind}))$ to
keep the radial field component $B_r$ constant at $\theta=\theta_{wind}$.
The initial $B_\varphi$ component is again zero throughout. In summary, we
now have the following set of parameters used in the simulation: the escape
speed $v_{esc}$, the polytropic index $\gamma$, the rotational
parameter $\zeta$, the field strength through $\beta$, and the extent of
the dead zone through $\theta_{wind}$. In addition, the mass flux $f_{mass}$
is used in the boundary condition of the poloidal momentum components.
Boundary conditions at the stellar surface are identical as above,
but now the dead zone has a zero mass flux, so that $f_{mass}(\theta)$.
Note that in a completely analoguous way, we could allow 
for a latitudinal dependence of the stellar rotation rate $\Omega_*(\theta)$, 
or the magnetic field strength $\beta(\theta)$. 

\begin{figure}
\begin{center}
\FIG{
\resizebox{\columnwidth}{!}{\includegraphics{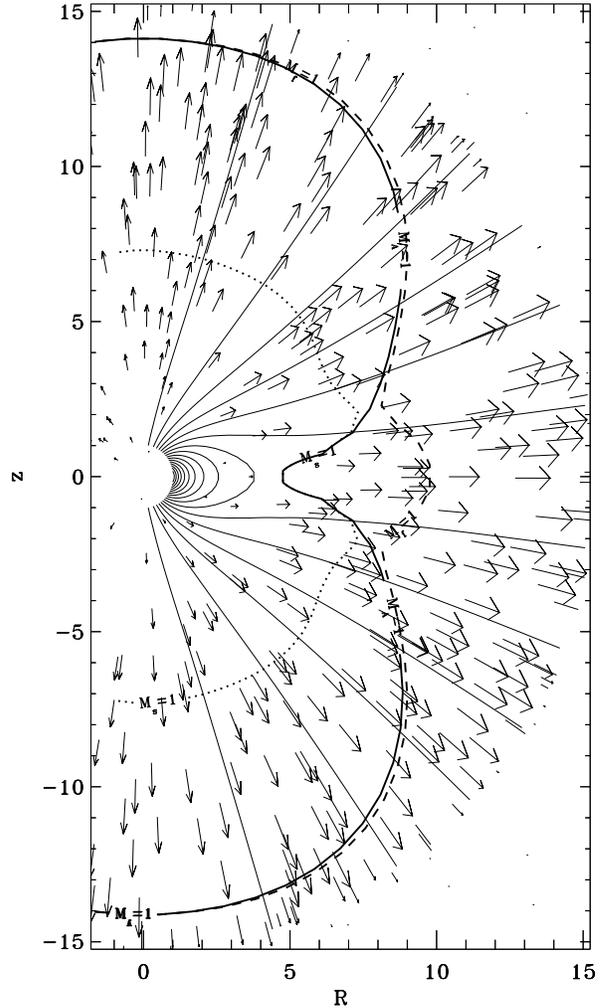}}
}
\vspace*{-0.5cm}
\caption{Axisymmetric magnetized wind containing a `wind' and a `dead' zone.
Shown are the
poloidal magnetic fieldlines and the poloidal flow field as vectors. Indicated
are the three 
critical surfaces where $M_s^p=1$ (dotted), $M_A^p=1$ (solid line),
and $M_f^p=1$ (dashed).}
\label{f-sakdead}
\end{center}
\end{figure}

This $t=0$ guess for an axisymmetric MHD wind is then time-advanced
to a stationary solution. Figure~\ref{f-sakdead} shows the final
stationary state, for
the parameter values $v_{esc}=3.3015c_{s*}$, $\gamma=1.13$,
a constant rotation 
rate corresponding to $\zeta=0.0156$, $\beta=3.69$, $\theta_{wind}=60^\circ$
and the mass flux in the wind zone set to the constant $f_{mass}=0.01377$, while
it is zero in the dead zone. These parameters are as in the WD solution
and the Sakurai wind presented earlier.
The initial field geometry has evolved to one where the open fieldlines are
draped around a distinct bipolar `dead' zone of limited radial extent and the
prescribed latitudinal range. The outflow nicely traces the field geometry
outside this dead zone. As seen from the figure, we have calculated
the full poloidal halfplane and imposed symmetry boundary conditions at north
and south pole. We used a polar grid of resolution $300\times 40$
of radial extent $[1,50]r_*$ with a radial grid accumulation at the
base. The north-south symmetry of the final solution is a firm
check of the numerics. The critical surfaces are also indicated in 
Fig.~\ref{f-sakdead} and they differ significantly from the monopolar
field solution shown in Fig.~\ref{f-polymhd2d}. Again, at the polar
regions, the Alfv\'{e}n and 
fast critical surface coincides. Now, the $B_{\varphi}$
also vanishes at the equator where conditions are such that
the slow and the Alfv\'en critical surfaces coincide. The $B_{\varphi}$
component changes sign when going from north to south, 
as the rigid rotation shears the initial,
purely poloidal bipolar magnetic field.
This is different from the Sakurai wind presented above, where the boundary
condition on $B_{\varphi}$ was taken symmetric about the equator.
Note how the equatorial
acceleration to super-Alfv\'enic velocities occurs very close
to the end of the dead zone. The critical surfaces are all displaced
inwards as compared to the monopolar case. 

\begin{figure}
\begin{center}
\FIG{
\hspace*{-2cm}
\resizebox{1.3\columnwidth}{!}{\includegraphics{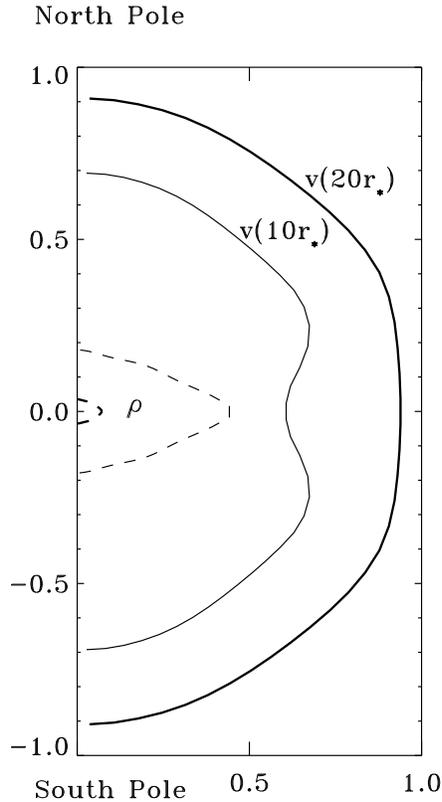}}
}
\vspace*{-0.5cm}
\caption{Polar plots of the scaled density (dashed) and velocity
(solid) for two fixed radial distances: 10 and 20 (thick lines)
stellar radii. The
`dead' zone has a clear influence on the latitudinal variation.}
\label{f-radcut}
\end{center}
\end{figure}

Figure~\ref{f-sakdead} shows that poloidal streamlines and fieldlines
are parallel. The $E_{\varphi}$ is below 3\%. In Fig.~\ref{f-radcut},
we show the latitudinal variation of the (scaled) density and the velocity 
at two fixed radial distances in a polar plot. The spacecraft Ulysses
and the on-board SWOOPS experiment
provided the solar community with detailed measurements of these quantities
for the solar wind (McComas et al.~\cite{swoops}). Qualitatively, the measured
poloidal density and velocity variation resembles the curves from 
Fig.~\ref{f-radcut}: the density is higher about the ecliptic and there is
a decrease in wind speed associated with the equatorial `dead' zone. 
However, our computational domain extended to 50 stellar radii, while
Ulysses measurements apply to larger radial distances. Note that we
could use observed solar differential rotation profiles, as well as
mass fluxes and magnetic field strengths, to obtain a better MHD proxy
of solar wind conditions. The extent of the solar coronal
active region belt suggests the use of a `dead' zone larger than modeled in
Fig.~\ref{f-sakdead}.

\section{Conclusions and outlook}\label{s-concl}

We obtained polytropic stellar winds as steady-state transonic 
outflows calculated with the Versatile Advection Code. 
We could relax an isothermal, spherically symmetric Parker wind,
to a polytropic wind model. Subsequently,
we included stellar rotation and a magnetic field, to arrive
at the well-known Weber-Davis solution. We used fully implicit
time stepping to converge to the steady-state solutions. The
correctness of these 1D wind solutions can be checked
{\it precisely}. 

We generalized to 2D axisymmetric, unmagnetized and magnetized
winds. Noteworthy is our prescription of the stellar boundary
conditions in terms of the prescribed mass flux $f_{mass}$ and
the way in which the parallelism of the flow and the fieldlines
in the poloidal plane is achieved. In Bogovalov (\cite{bogoval}), 
the stellar boundary
specified the normal magnetic field component and the density at the surface,
while keeping the velocity of the plasma on the stellar surface in the
rotating frame constant. Our approach differs markedly, since we impose the
mass flux and ensure the correct rotational
coupling of velocity and magnetic field. We refrain from fixing the
density, as the analytical treatment shows that the algebraic
Bernoulli equation together with the cross-field momentum
balance really determines the density profile and the magnetic flux function
concurrently, and should not be specified a priori.
In fact, we let the density and all magnetic field components adjust freely
at the base. This allows for the simultaneous and self-consistent
modeling of both open and closed fieldline regions, which is not
immediately possible when using the method of Sakurai (\cite{sakuraiAA}).
By an appropriate initialization of the time-marching
procedure used to get the steady-state solutions, we can find 
magnetized winds containing both a `wind' and a `dead' zone.

The method lends itself to investigate 
thermally and/or magneto-centrifu\-gal\-ly driven polytropic wind
solutions. One could derive angular momentum
loss rates used in studies of stellar rotational 
evolution (Keppens et al.~\cite{rotpaper}, Keppens~\cite{binary}).
However, our immediate interest is in the relaxation of the assumptions
inherent in our approach.

In this paper, we assume a polytropic equation of state throughout.
All solutions are smooth and
demonstrate a continuous acceleration from
subslow outflow at the stellar surface to superfast outflow at large
radial distances. 
Our polytropic assumption has to be relaxed to investigate the combined
coronal heating/solar wind problem within an MHD context. 
This involves adding the energy equation. We plan to study 
possible discontinuous transonic solutions containing shocks. We can
then address the puzzling paradox recently raised by
analytic investigations of translational symmetric and
axisymmetric transonic MHD flows (Goedbloed \& Lifschitz~\cite{hans1},
Lifschitz \& \linebreak Goedbloed~\cite{hans2}, Goedbloed et al.~\cite{eps98}). 
The generalized Grad-Shafranov equation describing
the cross-fieldline force balance has to be solved concurrently with
the algebraic condition expressed by the Bernoulli equation. 
Rigorous analysis of the generalized mixed-type Grad-Shafranov
partial differential equation, in combination with the algebraic 
Bernoulli equation, shows that only shocked solutions can be
realized whenever a limiting line appears within the domain of
hyperbolicity. 
Moreover, in Goedbloed \& Lifschitz (\cite{hans1}) and
Lifschitz \& Goedbloed (\cite{hans2}), it was pointed out that there are
forbidden flow regimes for certain
translationally symmetric, self-similar solutions of the MHD equations.
The Alfv\'en critical point is in those solutions situated within
a forbidden flow regime, which can only be crossed by shocks. 
It is of vital importance to understand what
ramifications this has on analytic and numerical studies of stellar
winds, or on accretion-type flows where
shocked solutions are rule rather than exception. Since the schemes
used in VAC are shock-capturing, we have all ingredients needed to
clarify this paradox.
Numerical studies of self-similar solutions as those discussed
in Trussoni et al. (\cite{tsinganos}) and Tsinganos et al. (\cite{tsinganos1})
are called for.
Combined analytic and numerical studies of such axisymmetric
steady-state flows have been initiated in Goedbloed et al. (\cite{eps98})
and in Ustyugova et al. (\cite{love}). 

After those paradoxes are resolved, we will be in a position to relax
the conditions of axisymmetry and stationarity. While several authors
have already initiated this daunting task (Gibson \& Low~\cite{gibson},
Guo \& Wu~\cite{guo}, Wu \& Dryer~\cite{wu}, Usmanov \& Dryer~\cite{usmanov}), 
we believe that an in-depth study of the subtleties involved with
the various restrictions mentioned is still warranted.

\begin{acknowledgements}
     The Versatile Advection Code was developed as part of the project on
     `Parallel Computational Magneto-Fluid Dynamics', funded by the
     Dutch Science Foundation (NWO)
     Priority Program on Massively Parallel Computing, and
     coordinated by JPG. Computer time on the Cray C90 
     was sponsored by the
     Dutch `Stichting Nationale Computerfaciliteiten' (NCF).
\end{acknowledgements}

\end{document}